\documentclass[sigconf]{acmart}
\usepackage{amsmath,graphicx,hyperref,booktabs,multirow,colortbl,subcaption,color,float,placeins}

\AtBeginDocument{%
  }

\setcopyright{cc}
\setcctype{by}
\copyrightyear{2026}
\acmYear{2026}
\acmConference[WWW '26] {Proceedings of the ACM Web Conference 2026}{April 13--17, 2026}{Dubai, United Arab Emirates.}
\acmBooktitle{Proceedings of the ACM Web Conference 2026 (WWW '26), April 13--17, 2026, Dubai, United Arab Emirates}
\acmISBN{979-8-4007-2307-0/2026/04}
\acmDOI{10.1145/3774904.3792339}

\begin{document}

\title{SDR-CIR: Semantic Debias Retrieval Framework for Training-Free Zero-Shot Composed Image Retrieval}


\author{Yi Sun}
\authornote{Both authors contributed equally to this research.}
\affiliation{%
  \institution{Wuhan University of Technology}
  \city{Wuhan}
  \country{China}}
\email{syi1005@whut.edu.cn}

\author{Jinyu Xu}
\authornotemark[1]
\affiliation{%
  \institution{Wuhan University of Technology}
  \city{Wuhan}
  \country{China}}
  \email{jinyxu@whut.edu.cn}

\author{Qing Xie}
\authornote{Corresponding author.}
\affiliation{%
 \institution{Wuhan University of Technology}
  \city{Wuhan}
  \country{China}}
  \email{felixxq@whut.edu.cn}

\author{Jiachen Li}
\affiliation{%
  \institution{Wuhan University of Technology}
  \city{Wuhan}
  \country{China}}
  \email{lijiachen@whut.edu.cn}

\author{Yanchun Ma}
\affiliation{%
  \institution{Wuhan Vocational College of Software and Engineering}
  \city{Wuhan}
  \country{China}}
\email{mayanchun@whvcse.edu.cn}

\author{Yongjian Liu}
\affiliation{%
  \institution{Wuhan University of Technology}
  \city{Wuhan}
  \country{China}}
\email{liuyj@whut.edu.cn}

\renewcommand{\shortauthors}{Yi Sun et al.}

\begin{abstract}
Composed Image Retrieval~(CIR) aims to retrieve a target image from a query composed of a reference image and modification text. Recent training-free zero-shot methods often employ Multimodal Large Language Models~(MLLMs) with Chain-of-Thought~(CoT) to compose a target image description for retrieval. However, due to the fuzzy matching nature of ZS-CIR, the generated description is prone to semantic bias relative to the target image.
We propose SDR-CIR, a training-free Semantic Debias Ranking method based on CoT reasoning. First, Selective CoT guides the MLLM to extract visual content relevant to the modification text during image understanding, thereby reducing visual noise at the source. 
We then introduce a Semantic Debias Ranking with two steps, Anchor and Debias, to mitigate semantic bias. 
In the Anchor step, we fuse reference image features with target description features to reinforce useful semantics and supplement omitted cues. 
In the Debias step, we explicitly model the visual semantic contribution of the reference image to the description and incorporate it into the similarity score as a penalty term. 
By supplementing omitted cues while suppressing redundancy, SDR-CIR mitigates semantic bias and improves retrieval performance. Experiments on three standard CIR benchmarks show that SDR-CIR achieves state-of-the-art results among one-stage methods while maintaining high efficiency. The code is publicly available at \url{https://github.com/suny105/SDR-CIR}.
\end{abstract}





\maketitle

\section{Introduction}

Composed Image Retrieval~(CIR)~\cite{vo2019composing,wen2023target,zhang2021heterogeneous} aims to retrieve a target image based on a reference image and a modification text. 
In recent years, CIR has been widely applied in e-commerce~\cite{du2023multi} and web search~\cite{jandial2022sac,psomas2024composed} due to its flexible query composition. 
However, CIR remains a challenging task as it requires a joint understanding of the visual content in the reference image and the semantic change conveyed by the modification text.
Conventional CIR methods~\cite{vo2019composing,liu2021image,baldrati2022effective,baldrati2022conditioned,ge2025llm} are typically trained on triplets comprising a reference image, a modification text, and a target image, but constructing such triplets is highly labor-intensive and limits the generalization capability of CIR. 
To mitigate this issue, Zero-Shot CIR~(ZS-CIR)~\cite{baldrati2023zero,saito2023pic2word,jang2024spherical,tang2024context} methods leverage the zero-shot capabilities of large-scale pre-trained visual-language models~(VLMs)~\cite{radford2021learning,li2022grounded} to perform retrieval without task-specific triplets. 

Existing ZS-CIR methods can be broadly classified into training-dependent and training-free methods~\cite{tang2025reason}. 
Training-dependent methods, such as textual inversion~\cite{baldrati2023zero,gu2024language,saito2023pic2word,cohen2022my,suo2024knowledge,lin2024fine,jang2024spherical}, learn a projection module from image-text pairs to map a reference image into pseudo-tokens, which are then composed with the modification text to perform retrieval. 
However, these methods still rely on additional training or fine-tuning on large-scale image-text pairs. 
In contrast, training-free methods~\cite{karthik2023vision,yang2024ldre,yang2024semantic,tang2025reason} directly utilize Large Language Models~(LLMs) or Multimodal Large Language Models~(MLLMs) to generate the composed query.
With the rapid development of large models such as GPT~\cite{openai_gpt41_docs_2025} and LLaVA~\cite{liu2023visual}, training-free ZS-CIR methods are gaining greater potential and broader application prospects.
Training-free ZS-CIR methods generally consist of a description generation process and a ranking process.
In the first process, LLMs or MLLMs are employed to generate a target image description that describes the target image by composing the reference image and the modification text. 
In the second process, the generated description is used as a text query to perform text-to-image retrieval based on similarity ranking.

\begin{figure}[tb]
    \centering
    \includegraphics[width=1\linewidth]{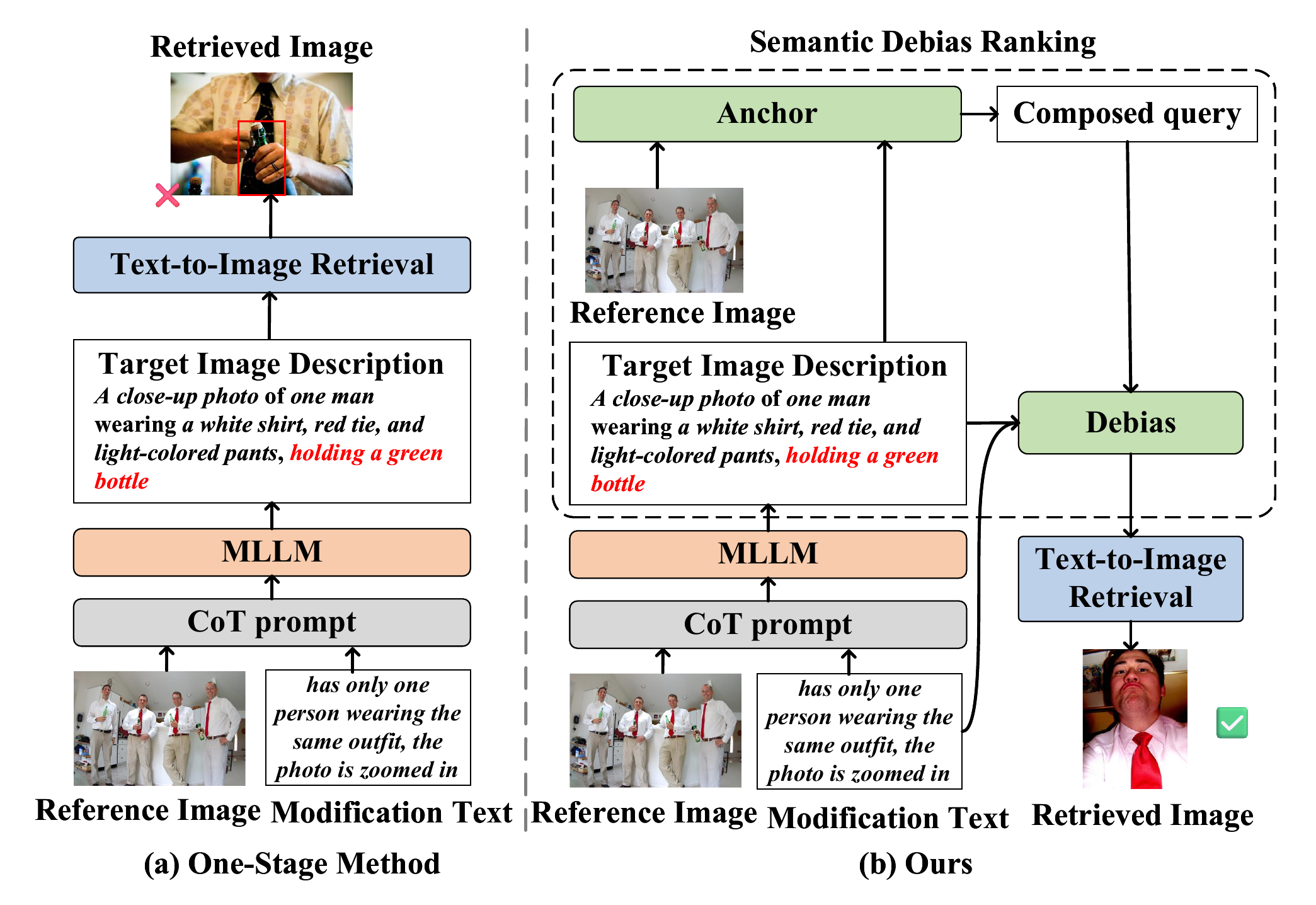}
    \caption{Comparison of existing one-stage method and our proposed method SDR-CIR.}
    \label{fig:1}
\end{figure}
Training-free ZS-CIR methods can be further categorized into two-stage methods and one-stage methods according to whether the reference image needs to be preprocessed using a captioning model during the description generation process~\cite{tang2025reason}. 
The two-stage methods~\cite{karthik2023vision,sun2023training,yang2024ldre,yang2024semantic,luo2025imagescope} first utilize a captioning model~\cite{li2022blip,li2023blip,dai2023instructblip} to obtain an image caption of the reference image, and then compose this caption with the modification text through LLMs to generate the target image description. 
However, since the reference image captioning is conducted independently of the modification text, the generated captions may omit key visual details relevant to the intended modification~\cite{Sun_2025_ICCV}. 
Consequently, LLMs struggle to infer accurate semantics under key visual information loss, which degrades the overall retrieval performance.
To address this limitation, one-stage methods leverage MLLMs to handle the reference image and the modification text simultaneously through elaborately designed prompts, thereby ensuring that both visual details and textual intent are jointly considered during target image description generation. 
Despite this progress, most existing one-stage methods that directly employ MLLMs for CIR still face two challenges:

\textbf{(1) Visual noise in the description generation process. } 
In ZS-CIR, the multimodal query is inherently uncertain and the reference image includes redundant visual semantics.
Most existing one-stage methods \cite{tang2025reason} adopt a Chain-of-Thought~(CoT)~\cite{gao2024cantor,wei2022chain}, strategy that first extracts nearly all visual information from the reference image and then filters irrelevant details according to the modification text. 
However, this manner inherently lacks semantic selectivity during the initial extraction. 
Since visual extraction is driven by the overall image content rather than the modified intent, irrelevant visual information can interfere with the subsequent filtering. 
Therefore, the generated target image description may introduce noise, ultimately degrading retrieval performance.

\textbf{(2) Semantic bias in the ranking process. }
ZS-CIR can be regarded as a fuzzy matching task~\cite{bordogna1993fuzzy,chen2002region}, since the correspondence between the reference and target images is not strictly defined by the modification text. 
In practice, the target image may differ from the reference image in additional, implicit semantic aspects beyond the specified modification. 
Consequently, the generated target image description only partially corresponds to the target image, resulting in inevitable semantic bias in the description with respect to the target image.
For example, as shown in Figure~\ref{fig:1}, "a green bottle" is unrelated to the target image, but it is a person-related attribute in the reference image. Since the person is the primary subject in the image, MLLMs may treat it as a useful cue and include it in the target image description.
This bias includes the redundancy bias and the omission bias: the target image description either contains redundant details or omits useful cues to the target image.
Consequently, such semantic bias can misguide the ranking process, as irrelevant or inaccurate concepts in the generated description may falsely increase the similarity scores of non-target candidates while reducing the relevance of the true target image.



To address these challenges, we propose a novel training-free \textbf{S}emantic \textbf{D}ebias \textbf{R}etrieval framework for zero-shot \textbf{C}omposed \textbf{I}mage \textbf{R}etrieval~(\textbf{SDR-CIR}).
In the generation process, unlike previous one-stage CoT-based methods that extract all visual content before filtering, we design a \textbf{Selective CoT} that guides MLLMs to selectively extract visual content relevant to the modification text, thereby mitigating visual noise during reference image understanding.
Conceptually, Selective CoT serves as a reasoning-control mechanism that constrains the inference trajectory of the MLLM to the semantic scope relevant to the modification text, which helps reduce redundant visual reasoning, resulting in a generated target image description that is more semantically aligned with the intended target image.

To further alleviate the semantic bias introduced by visual semantics from the reference image, we design a \textbf{Semantic Debias Ranking} module in the ranking process.
Unlike a prior approach~\cite{yang2024semantic} that relies on semantic editing increments to alleviate the semantic bias, since the bias between the target description and the target image is caused by the fuzzy matching between the reference and target images, the semantic bias in the description essentially stems from the reference image.
Therefore, we explicitly model and correct the visual semantic contribution of the reference image that may incorrectly dominate the similarity computation, thereby mitigating the semantic bias.
Specifically, this module consists of two steps: the \textit{Anchor} step and the \textit{Debias} step.
In the \textit{Anchor} step, we establish a semantic anchor by fusing the reference image and generated description features to reinforce the useful visual semantics and supplement omitted visual cues.
In the \textit{Debias} step, we then adjust the final similarity by penalizing the biased visual semantic contribution.
This design enables our approach to supplement omitted visual cues from the reference image while suppressing redundant semantics, achieving more accurate and balanced ranking results.


In summary, our main contributions are as follows:
\begin{itemize}

\item{We propose a training-free \textbf{S}emantic \textbf{D}ebias \textbf{R}etrieval framework for zero-shot \textbf{C}omposed \textbf{I}mage \textbf{R}etrieval~(\textbf{SDR-CIR}). This framework addresses both visual noise in the description generation process and semantic bias in the ranking process in ZS-CIR.}

\item{We design a Selective CoT that guides multimodal large language models~(MLLMs) to selectively extract visual semantics relevant to the modification text, thereby mitigating visual noise during reference image understanding and improving the target image description generation.}

\item{We introduce a Semantic Debias Ranking module with an Anchor–then–Debias strategy to correct biased visual semantic contributions and achieve balanced retrieval performance.}

\item{Experiments on three standard benchmarks show consistent improvements over previous methods.}
\end{itemize}

\section{Related Work}
\subsection{Zero-Shot Composed Image Retrieval}
Zero-Shot Composed Image Retrieval~(ZS-CIR)~\cite{baldrati2023zero,karthik2023vision,tang2025reason,gu2024language,yang2024ldre,yang2024semantic,saito2023pic2word,Sun_2025_ICCV,tang2024context} leverages pre-trained VLMs to avoid CIR-style triplet supervision.
These methods primarily focus on how to integrate a reference image and a modification text to form a composed query for retrieval.
Training-dependent ZS-CIR methods~(e.g., Pic2Word~\cite{saito2023pic2word}, SEARLE~\cite{baldrati2023zero}) learn a projection module to map the reference image into pseudo-tokens and then compose them with the modification text, but they still require additional training on large-scale image–text pairs, motivating the development of training-free pipelines.
Two-stage training-free methods~(e.g., CIReVL~\cite{karthik2023vision}, LDRE~\cite{yang2024ldre}, SEIZE~\cite{yang2024semantic}) first convert the reference image into a caption and then compose this caption with the modification text through LLM inference.
However, since the reference image captioning is performed independently of the modification text, these methods may omit key visual details relevant to the intended modification, leading to inaccurate reasoning. 
In contrast, one-stage training-free methods, such as OSrCIR~\cite{tang2025reason} and CoTMR~\cite{Sun_2025_ICCV}, jointly input the reference image and the modification text into an MLLM to ensure that visual details and modification intent are jointly considered.
Many methods perform retrieval and ranking using only generated descriptions~\cite{karthik2023vision,tang2025reason}, or by combining generated descriptions with multi-scale objects~\cite{Sun_2025_ICCV}.
Nevertheless, since ZS-CIR is essentially a fuzzy matching task, the generated target image description inevitably exhibits semantic bias with respect to the target image.
To address this issue, we propose a Semantic Debias Ranking module that alleviates this semantic bias by explicitly modeling and penalizing the visual semantic contribution of the reference image to the description in one-stage methods. 

\subsection{Chain-of-Thought in ZS-CIR}
Recent studies have demonstrated that Chain-of-Thought~(CoT) prompts~\cite{wei2022chain} can substantially enhance the reasoning and comprehension capabilities of LLMs or MLLMs. 
The key idea of CoT is to design task-specific prompts that encourage the model to generate intermediate reasoning steps, which has proven effective across various domains~\cite{zhang2022automatic,miao2024chain,mitra2024compositional,zhang2023multimodal,suzgun2022challenging,feng2023towards,lyu2023faithful}.
In ZS-CIR task, several one-stage methods~\cite{tang2025reason,Sun_2025_ICCV} have recently explored incorporating CoT to improve the reasoning capabilities of MLLMs. 
For instance, OSrCIR~\cite{tang2025reason} introduces a reflective CoT to help MLLMs understand and implement modifications. 
CoTMR~\cite{Sun_2025_ICCV} leverages CoT reasoning to enable MLLMs to apply modifications step by step. 
These approaches effectively enhance the reasoning ability of MLLMs in integrating the visual and textual modalities for retrieval. 
However, these methods ignore the redundancy of the reference image. 
Existing ZS-CIR methods based on CoT require MLLMs to extract nearly all visual information before filtering irrelevant details, which inherently lacks semantic selectivity. 
Therefore, irrelevant or redundant visual content may be included in the generated target image description, introducing visual noise that affects retrieval performance.
To mitigate this problem, we propose a Selective CoT, which guides the MLLM to selectively extract visual content relevant to the modification text during reference image understanding.

\section{Methodology}
\begin{figure*}[t]
    \centering
    \includegraphics[width=0.9\linewidth,trim=5 10 5 10,clip]{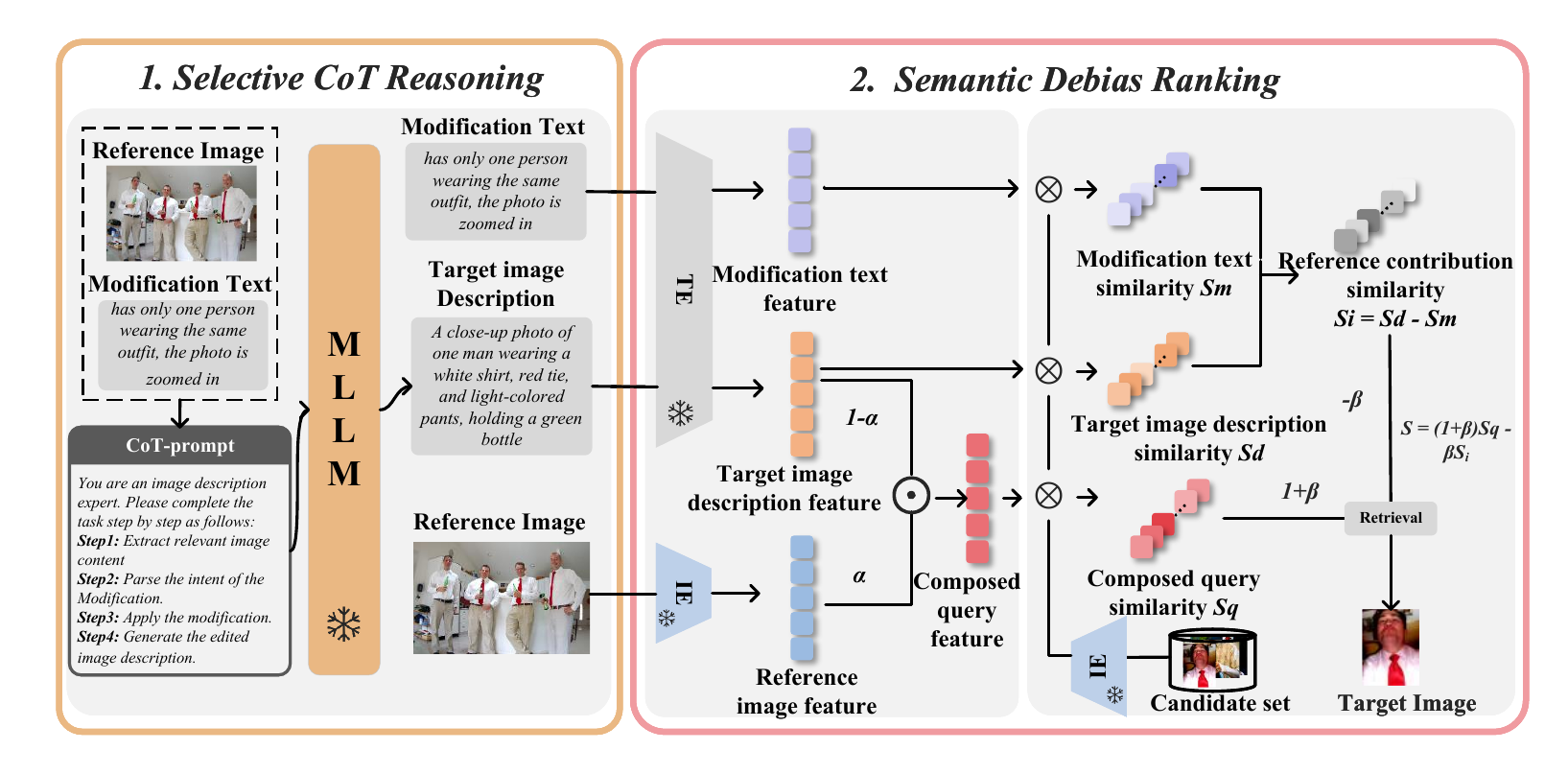}
    \caption{Overview of our SDR-CIR framework. (1) The Selective CoT prompt instructs the MLLM to extract visual content relevant to modification guided by the modification text. (2) Semantic Debias Ranking: we fuse reference image and target image description feature as the composed query to anchor the useful visual semantics and then represent the similarity between the visual semantic contribution and candidate images as a penalty term to debias. }
    \label{fig:2}
\end{figure*}

In this section, we introduce SDR-CIR, a training-free ZS-CIR method that mitigates semantic bias~(Fig.~\ref{fig:2}). Section~\ref{3.1} provides an overview of our approach. Section~\ref{3.2} presents the design of our Selective CoT prompt, and Section~\ref{3.3} introduces the Semantic Debias Ranking module, which alleviates the semantic bias.

\subsection{Overview}
\label{3.1}
Figure~\ref{fig:2} illustrates the overall framework.
We first employ a Selective CoT to reduce visual noise in the generation process by guiding the MLLM to extract modification-relevant visual elements.
We then perform retrieval with a Semantic Debias Ranking~(SDR) module to mitigate semantic bias in ranking, where the \textit{Anchor} step constructs a robust composed query and the \textit{Debias} step suppresses non-target candidates by penalizing reference-induced semantic bias, thereby improving retrieval performance.

\subsection{Selective CoT Reasoning}
\label{3.2}
In the CIR task, a user query consists of a reference image and a modification text. The reference image provides the visual content for retrieving the target image.
In fact, the visual content provided by the reference image tends to include redundant information.
However, existing one-stage methods ignore this fact and employ CoT prompts to guide MLLMs to extract nearly all visual information when understanding the reference image, which can introduce noisy details into the generated description and then affect retrieval performance. 
To mitigate this problem, we design a selective CoT prompt: guided by the modification text, the MLLM selectively extracts visual content relevant to the modification during reference image understanding. 

Similar to existing CoT prompts, our prompt adopts a four-stage structure: (1) reference image understanding, (2) modification text understanding, (3) applying modification, and (4) target image description generation. 
The key difference is that we introduce the guidance of the modification text during the image understanding to extract visual content relevant to the modification.

\begin{figure}[tb]
\includegraphics[width=0.9\linewidth, trim=10 20 10 20,clip]{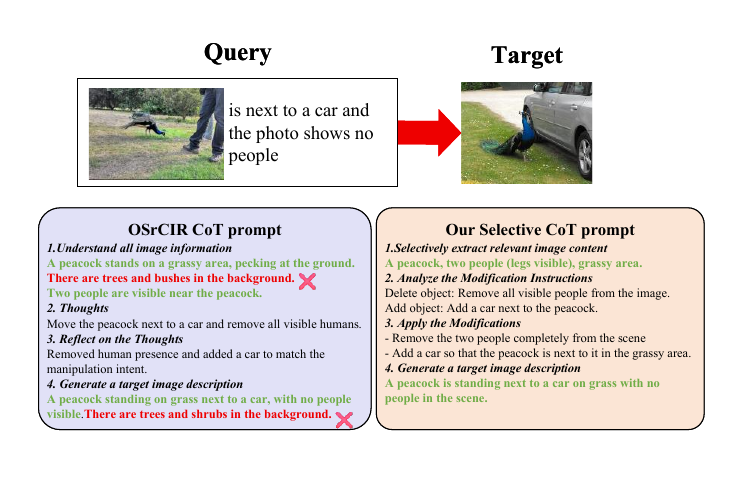}
\caption{Comparison on CoT prompt between OSrCIR and ours.}
\label{fig:3}
\end{figure}

Specifically, in the reference image understanding, we first parse the modification text to infer explicit modified targets~(directly specified), and infer implicit modified targets~(implied by the modification text).
Then we understand the reference image and selectively extract the visual content relevant to the modification text guided by these targets.
For example, in Figure~\ref{fig:3}, the visual content relevant to the modification text includes: a peacock, two people~(legs visible), and a grassy area, rather than the background. 
This selective process helps reduce visual noise in the target image description during the generation process.

Then the MLLM infers the modification intent from the modification text, applies modifications step by step to update the extracted visual content, and finally obtains the target image description.
\subsection{Semantic Debias Ranking}
\label{3.3}
To alleviate the semantic bias introduced by the biased visual semantic contribution from the reference image, we propose a Semantic Debias Ranking module comprising two steps: the \textit{Anchor} step and the \textit{Debias} step. 
The \textit{Anchor} step reinforces related useful semantics and supplements omitted cues by fusing reference image features with the target image description features. Then the \textit{Debias} step penalizes the biased visual semantic contribution of the reference image to degrade the ranking of non-target candidate images, thereby debiasing and improving retrieval performance.
\subsubsection{Anchor: reinforcing and supplementing information}
The semantic bias between the target image description and the target image arises from the fuzzy correspondence between the reference image and the target image.
As a result, the bias in the description primarily stems from the reference image.
In practice, biased descriptions exhibit two types of bias: redundancy bias and omission bias. 
In redundant cases, the generated target description retains details irrelevant to the target image from the reference image.
In omitted cases, it omits key cues from the reference image. 
These biases reflect excessive or insufficient representation of the reference image content. 
Notably, the visual semantic contribution of the reference image also contains the useful visual semantics to the target image.
Directly suppressing the visual contribution may lead to the loss of useful semantics. 
Therefore, we leverage the reference image feature and the target image description feature to construct a robust composed query feature, reinforcing useful visual semantics and supplementing omitted key cues in the description.

Specifically, we establish a semantic anchor by fusing the reference image feature and the target image description feature. 
We employ the CLIP text encoder to encode the description $T_d$ into feature $F_d$.
The image $I_t^i$ in the candidate image set $D$ and reference image $I_r$ are encoded into $F(I_t^i)$ and $F_r$ through the CLIP image encoder, ensuring that they reside in the same feature space. 
Subsequently, we fuse reference image feature $F_r$ with target image description feature $F_d$ to obtain the composed query feature $F_q$.
\begin{equation}
    F_q = (1-\alpha)F_d +\alpha F_r,
\end{equation}
where $\alpha$ is the weight assigned to the reference image feature.

Because useful visual semantics dominate~(such as "a white shirt, red tie and light-colored", as shown in the Figure~\ref{fig:2}) the reference image, for redundant descriptions, adding image features increases the relative weight of useful information and prevents it from being weakened when penalizing the visual semantic contribution. 
For incomplete descriptions, image features can supplement omitted cues. 
This \textit{Anchor} step obtains a robust composed query that keeps useful semantics salient while supplementing omitted cues.

\subsubsection{Debias: penalizing the visual semantic contribution}
As the reference image is the main source of semantic bias in the description, we can mitigate the semantic bias by expressing the reference image’s visual semantic contribution to the target image description.
Inspired by SEIZE~\cite{yang2024semantic}, we represent the visual semantic contribution in the description from a similarity perspective. 
The target image description is generated through the joint influence of the reference image and the modification text, and the modification text usually contains correct information. 
As a result, the reference image’s semantic contribution can be approximated as the difference between the target image description and the contribution from the modification text. 

Specifically, we compute cosine similarities $S_q$, $S_d$, $S_m$ between the composed query feature $F_q$, target image description feature $F_d$, the modification text feature $F_m$, respectively, and image feature $F(I_t^i)$ from the candidate image set $D$:
\begin{equation}
    [S_q,S_d,S_m]= sim([F_q,F_d,F_m],F(I_t^i)),  \forall I_t^i \in D,
\end{equation}

Therefore, we can obtain the cosine similarity $S_i$ between the candidate image and the visual semantic contribution of the reference image by calculating the difference between $S_d$ and $S_m$. 
\begin{equation}
    S_i = S_d - S_m,
\end{equation}

For redundant descriptions, $S_i$ primarily represents the similarity between the candidate image and the visual semantic contribution that includes redundant visual information. 
For incomplete descriptions, $S_i$ primarily represents the similarity between the candidate image and the visual semantic contribution that omits key cues. 
When $S_i$ has a high value, it generally means that the candidate image is consistent with redundant or incomplete visual semantic contribution. 
Such a candidate is more likely to be a non-target image. 
Therefore, we include $S_i$ as a penalty term in the final similarity to penalize candidate images that are likely to be non-target.

We subtract the similarity $S_i$ of the visual semantic contribution from the composed query similarity score $S_q$ to obtain the final similarity $S_f$. 
\begin{equation}
    S_f = (1+\beta )S_q-\beta S_i,
\end{equation}
where $\beta$ is the weight that suppresses the contribution from the reference image. Meanwhile, we set the weight of $S_q$ to $1+\beta$ to balance the score.

Finally, the final similarity score $S_f$ is used for retrieval. 

\section{Experiments}

\begin{table*}[]
\caption{\textbf{Comparison results on the CIRCO and CIRR test sets.} The highest performance values are shown in \textbf{bold}, and the second-highest are \underline{underlined}. "2S" indicates two-stage methods. "1S" indicates one-stage methods. "*" indicates experiments with GPT-4.1 as the MLLM and using CLIP weights from~\cite{openclip_zenodo_5143773} without using context examples in CoT. }
\scalebox{0.8}{
\begin{tabular}{@{}c|ccc|cccc||ccccccc@{}}
\toprule
\multicolumn{3}{c}{\textbf{CIRCO+CIRR→}} & &
  \multicolumn{4}{c||}{\textbf{CIRCO}} &
  \multicolumn{7}{c}{\textbf{CIRR}} \\ 
  \midrule
  \multirow{2}{*}{Backbone} & \multirow{2}{*}{Method} & \multirow{2}{*}{Training-free} & \multirow{2}{*}{Type} &
  \multicolumn{4}{c||}{mAP@k} &
  \multicolumn{4}{c|}{Recall@k} &
  \multicolumn{3}{c}{Recall$_{\text{sub@}}$k} \\
   &
   &
    & &
  k=5 &
  k=10 &
  k=25 &
  k=50 &
  k=1 &
  k=5 &
  k=10 &
  \multicolumn{1}{c|}{k=50} &
  k=1 &
  k=2 &
  k=3 \\ \midrule
\multicolumn{1}{c}{} &
  \multicolumn{1}{|c}{SEARLE~\cite{baldrati2023zero}} &
  ×  & - &
  9.35 &
  9.94 &
  11.13 &
  11.84 &
  24.00 &
  53.42 &
  66.82 &
  \multicolumn{1}{c|}{89.78} &
  54.89 &
  76.60 &
  88.19 \\
  \rowcolor{gray!20}
\multicolumn{1}{c}{\cellcolor{white}{}} &
  \multicolumn{1}{|c}{CIReVL~\cite{karthik2023vision}} &
  \ensuremath{\surd}  & 2S &
  14.95 &
  15.42 &
  17.00 &
  17.82 &
  23.94 &
  52.51 &
  66.00 &
  \multicolumn{1}{c|}{86.95} &
  60.17 &
  80.05 &
  90.19 \\
  \rowcolor{gray!20}
\multicolumn{1}{c}{\cellcolor{white}{}} &
  \multicolumn{1}{|c}{ LDRE~\cite{yang2024ldre}} &
  \ensuremath{\surd}  & 2S &
  17.96 &
  18.32 &
  20.21 &
  21.11 &
  25.69 &
  55.13 &
  69.04 &
  \multicolumn{1}{c|}{89.90} &
  60.53 &
  80.65 &
  90.70 \\
  \rowcolor{gray!20}
\multicolumn{1}{c}{\cellcolor{white}{}} &
  \multicolumn{1}{|c}{SEIZE~\cite{yang2024semantic}} &
  \ensuremath{\surd}  & 2S&
  19.04 &
  19.64 &
  21.55 &
  22.49 &
  27.47 &
  57.42 &
  70.17 &
  \multicolumn{1}{c|}{-} &
  65.59 &
  84.48 &
  92.77 \\
\multicolumn{1}{c}{} &
  \multicolumn{1}{|c}{OSrCIR*~\cite{tang2025reason}} &
  \ensuremath{\surd}  & 1S &
  16.85&
  17.39&
  19.15&
  20.01&
  28.07&
  57.95&
  69.71&
  \multicolumn{1}{c|}{88.94} &
  62.31&
  81.18&
  91.04\\
\multicolumn{1}{c}{} &
  \multicolumn{1}{|c}{CoTMR*~\cite{Sun_2025_ICCV}} &
  \ensuremath{\surd}  & 1S &
  \underline{21.16}&
  \underline{21.77}&
  \underline{23.71}&
  \underline{24.70}&
  \underline{30.12}&
  \underline{60.19}&
  \underline{71.71}&
  \multicolumn{1}{c|}{\underline{90.34}} &
  \underline{67.11}&
  \underline{85.13}&
  \underline{93.64}\\
\multicolumn{1}{c}{\multirow{-7}{*}{ViT-B/32}} &
  \multicolumn{1}{|c}{\textbf{SDR-CIR}} &
  \ensuremath{\surd}  & 1S &
  \textbf{23.78}&
  \textbf{24.43}&
  \textbf{26.58}&
  \textbf{27.50}&
  \textbf{34.48}&
  \textbf{65.74}&
  \textbf{76.87}&
  \multicolumn{1}{c|}{\textbf{93.06}} &
  \textbf{69.90}&
  \textbf{87.04}&
  \textbf{94.48}\\ \midrule
\multicolumn{1}{c}{} &
  \multicolumn{1}{|c}{Pic2Word~\cite{saito2023pic2word}} &
  ×  &- &
  8.72 &
  9.51 &
  10.64 &
  11.29 &
  23.90 &
  51.70 &
  65.30 &
  \multicolumn{1}{c|}{87.80} &
  - &
  - &
  - \\
\multicolumn{1}{c}{} &
  \multicolumn{1}{|c}{SEARLE~\cite{baldrati2023zero}} &
  ×  &- &
  11.68 &
  12.73 &
  14.33 &
  15.12 &
  24.24 &
  52.48 &
  66.29 &
  \multicolumn{1}{c|}{88.84} &
  53.76 &
  75.01 &
  88.19 \\
\multicolumn{1}{c}{} &
  \multicolumn{1}{|c}{MLLM-I2W~\cite{bao2025mllm}} &
  ×  &- &
  - &
  - &
  - &
  - &
  28.30 &
  57.90 &
  70.20 &
  \multicolumn{1}{c|}{93.90} &
  - &
  - &
  - \\
  \rowcolor{gray!20}
\multicolumn{1}{c}{\cellcolor{white}{}} &
  \multicolumn{1}{|c}{CIReVL~\cite{karthik2023vision}} &
  \ensuremath{\surd}  & 2S &
  18.57 &
  19.01 &
  20.89 &
  21.80 &
  24.55 &
  52.31 &
  64.92 &
  \multicolumn{1}{c|}{86.34} &
  59.54 &
  79.88 &
  89.69 \\
  \rowcolor{gray!20}
\multicolumn{1}{c}{\cellcolor{white}{}} &
  \multicolumn{1}{|c}{LDRE~\cite{yang2024ldre}} &
  \ensuremath{\surd}  & 2S &
  23.25 &
  24.03 &
  26.44 &
  27.50 &
  26.53 &
  55.57 &
  67.54 &
  \multicolumn{1}{c|}{88.50} &
  60.43 &
  80.31 &
  89.90 \\
  \rowcolor{gray!20}
\multicolumn{1}{c}{\cellcolor{white}{}} &
  \multicolumn{1}{|c}{SEIZE~\cite{yang2024semantic}} &
  \ensuremath{\surd}  & 2S &
  24.98 &
  25.82 &
  28.24 &
  29.35 &
  28.65 &
  57.16 &
  69.23 &
  \multicolumn{1}{c|}{-} &
  66.22 &
  84.05 &
  92.34 \\
\multicolumn{1}{c}{} &
  \multicolumn{1}{|c}{OSrCIR*~\cite{tang2025reason}} &
  \ensuremath{\surd}  & 1S &
  21.83 &
  22.46 &
  24.49 &
  25.44 &
  30.63 &
  60.34 &
  72.00 &
  \multicolumn{1}{c|}{89.86} &
  64.31 &
  82.29 &
  91.35 \\
\multicolumn{1}{c}{} &
  \multicolumn{1}{|c}{CoTMR*~\cite{Sun_2025_ICCV}} &
  \ensuremath{\surd}  & 1S &
  \underline{26.52} &
  \underline{27.13}&
  \underline{29.51}&
  \underline{30.56}&
  \underline{33.54}&
  \underline{63.25}&
  \underline{74.63}&
  \multicolumn{1}{c|}{\underline{91.08}} &
  \underline{69.88}&
  \underline{86.53}&
  \underline{94.19}\\
\multicolumn{1}{c}{\multirow{-8}{*}{ViT-L/14}} &
  \multicolumn{1}{|c}{\textbf{SDR-CIR}} &
  \ensuremath{\surd}  & 1S &
  \textbf{30.91}&
  \textbf{31.50}&
  \textbf{34.03}&
  \textbf{35.08}&
  \textbf{37.61} &
  \textbf{67.71} &
  \textbf{79.13} &
  \multicolumn{1}{c|}{\textbf{93.81}} &
  \textbf{71.90} &
  \textbf{88.39} &
  \textbf{94.63} \\ \midrule
  \rowcolor{gray!20}
\multicolumn{1}{c}{\cellcolor{white}{}} &
  \multicolumn{1}{|c}{CIReVL~\cite{karthik2023vision}} &
  \ensuremath{\surd}  & 2S &
  26.77 &
  27.59 &
  29.96 &
  31.03 &
  34.65 &
  64.29 &
  75.06 &
  \multicolumn{1}{c|}{91.66} &
  67.95 &
  84.87 &
  93.21 \\
  \rowcolor{gray!20}
\multicolumn{1}{c}{\cellcolor{white}{}} &
  \multicolumn{1}{|c}{LDRE~\cite{yang2024ldre}} &
  \ensuremath{\surd}  & 2S &
  31.12 &
  32.24 &
  34.95 &
  36.03 &
  36.15 &
  66.39 &
  77.25 &
  \multicolumn{1}{c|}{93.95} &
  68.82 &
  85.66 &
  93.76 \\
  \rowcolor{gray!20}
\multicolumn{1}{c}{\cellcolor{white}{}} &
  \multicolumn{1}{|c}{SEIZE~\cite{yang2024semantic}} &
  \ensuremath{\surd}  & 2S &
  \underline{32.46} &
  \underline{33.77} &
  \underline{36.46} &
  \underline{37.55} &
  \underline{38.87} &
  \underline{69.42} &
  \underline{79.42} &
  \multicolumn{1}{c|}{-} &
  \textbf{74.15} &
  \textbf{89.23} &
  \textbf{95.71} \\
\multicolumn{1}{c}{} &
  \multicolumn{1}{|c}{OSrCIR*~\cite{tang2025reason}} &
  \ensuremath{\surd}  & 1S &
  24.73 &
  25.99 &
  28.09 &
  29.19 &
  33.52 &
  61.66 &
  73.30 &
  \multicolumn{1}{c|}{90.17} &
  65.42 &
  82.75 &
  91.54 \\
\multicolumn{1}{c}{} &
  \multicolumn{1}{|c}{CoTMR*~\cite{Sun_2025_ICCV}} &
  \ensuremath{\surd}  & 1S &
  29.59&
  30.74&
  33.37&
  34.44&
  35.93&
  65.11&
  75.33&
  \multicolumn{1}{c|}{91.45} &
  70.82&
  87.16&
  94.48\\
\multicolumn{1}{c}{\multirow{-6}{*}{ViT-G/14}} &
  \multicolumn{1}{|c}{\textbf{SDR-CIR}} &
  \ensuremath{\surd}  & 1S &
  \textbf{33.05}&
  \textbf{34.50}&
  \textbf{37.21}&
  \textbf{38.42}&
  \textbf{40.17} &
  \textbf{69.76} &
  \textbf{79.88} &
  \multicolumn{1}{c|}{\textbf{94.00}} &
  \underline{73.30} &
  \underline{88.89} &
  \underline{94.99} \\ \bottomrule
\end{tabular}
}

\label{tab:1} 
\end{table*}

\subsection{Experimental Setup}
\textbf{Datasets.} 
We conduct experiments on three widely used CIR datasets: CIRR~\cite{liu2021image}, CIRCO~\cite{baldrati2023zero}, and FashionIQ~\cite{wu2021fashion}. 
\textbf{CIRR} is the first open-domain dataset for CIR and contains 36,554 queries, each paired with a single target image. 
\textbf{CIRCO}, built from real images in COCO 2017 unlabeled set~\cite{lin2014microsoft}, includes 800 test queries and 220 validation queries. 
Each query is associated with multiple ground truths, averaging 4.53 per query. 
\textbf{FashionIQ} focuses on the fashion domain and covers three categories—shirt, dress, and toptee. 
It comprises 30,135 query triplets and 77,683 candidate images, with each query linked to a single target image. 
These datasets span different types of modification needs in CIR: CIRR and CIRCO emphasize object-level changes~(e.g., adding, removing, or replacing objects), while FashionIQ emphasizes attribute-level modifications~(e.g., clothing color and style). 

\noindent \textbf{Evaluation Metrics.} 
We adopt Recall@k~(R@k) for CIRR and FashionIQ, while mean average precision~(mAP@k) is used for CIRCO. 
In addition, we report Recall$_{\text{sub@k}}$ on the CIRR subset, which contains samples that are highly similar to the target image, to provide a more rigorous evaluation.

\noindent \textbf{Baselines.} 
We compare our method with several common ZS-CIR methods. 
For training-dependent ZS-CIR methods, we select: 
1) \textbf{Pic2Word}~\cite{saito2023pic2word}: maps the reference image feature into pseudo-tokens. 
2) \textbf{SEARLE}~\cite{baldrati2023zero}: combines the pseudo-word token with the caption produced by GPT. 
3) \textbf{MLLM-I2W}~\cite{bao2025mllm}: maps reference image features to pseudo-word tokens with MLLM contextual prompts. 
For training-free ZS-CIR methods, we compare one-stage and two-stage methods. 
For two-stage methods: 
1) \textbf{CIReVL}~\cite{karthik2023vision}: generates target image description using a two-stage process of BLIP-2~\cite{li2023blip} and an LLM. 
2) \textbf{LDRE}~\cite{yang2024ldre}: generates and aggregates multiple target image descriptions.
3) \textbf{SEIZE}~\cite{yang2024semantic}: generates and aggregates multiple target image descriptions and performs semantic editing search. 
For one-stage methods: 
1) \textbf{OSrCIR}~\cite{tang2025reason}: generates target image description using a MLLM based on CoT and performs direct search. 
2) \textbf{CoTMR}~\cite{Sun_2025_ICCV}: generates target image description using a MLLM based on CoT and performs multi-scale inference.

\noindent \textbf{Implementation Details.} 
We employ GPT-4.1~\cite{openai_gpt41_docs_2025} as the main MLLM for target image description generation, and use Qwen2.5-VL-72B~\cite{bai2025qwen2} and GPT-4o-mini~\cite{openai_gpt4omini_docs} for ablation analysis. 
For the retrieval encoder, we adopt three variants of the pre-trained CLIP model from OpenCLIP~\cite{openclip_zenodo_5143773}: ViT-B/32, ViT-L/14, and ViT-G/14. 
The scaling factors $\alpha$ and $\beta$ are set to 0.05 and 0.5 for CIRR dataset, 0.15 and 0.35 for CIRCO dataset, 0.2 and 0.4 for FashionIQ dataset, respectively. 
All experiments are implemented in PyTorch~\cite{paszke2019pytorch} and conducted on a single NVIDIA RTX 3090 GPU.

\subsection{ZS-CIR Benchmark Comparisons}
Our method follows a one-stage pipeline, and therefore other one-stage approaches are adopted as the main baselines for comparison. 
For a fair comparison, we employ the same MLLM~(GPT-4.1) across all one-stage baselines.

\textbf{CIRCO:}
The left part of Table~\ref{tab:1} presents the experimental results on the CIRCO test dataset. 
We make the following observations: 
(1) On CIRCO, where each query has multiple ground-truth images, an overly precise description may align with only a single instance, thereby narrowing the retrieval scope.
SDR-CIR alleviates this issue via semantic debiasing, suppressing reference-induced irrelevant details in the description to improve coverage of valid targets, achieving the best performance among all compared baselines.
(2) Compared with our direct baselines OSrCIR and CoTMR, using ViT-L/14 as an example, our SDR-CIR improves mAP@5 by 9.08\% and 4.39\%, respectively. Even without using in-context examples, our method still has significant effects while maintaining a friendly computational cost.

\textbf{CIRR:}
The right part of Table~\ref{tab:1} presents the experimental results on the CIRR test dataset and its subsets. 
We make the following observations: 
(1) CIRR dataset has many false negatives, and retrieval is primarily dominated by the modification text. 
Therefore, the content can highly affect the retrieval performance if the description exists the redundant reference image content. 
In this case, SDR-CIR can significantly suppress the interference of redundant information from the reference image, achieving the best performance among all baselines. 
For example, on the ViT-L/14 backbone, compared with the best training-free method CoTMR, our method improves Recall@1 by 4.07\%. 
(2) In the CIRR subset, SDR-CIR outperforms all one-stage baselines. 
For example, on ViT-G/14, compared to the best one-stage method, CoTMR, our method achieves a 2.48\% improvement in Recall$_{\text{sub@}}$1, but it still trails the two-stage SEIZE. 
The gap likely stems from the subset’s need for finer discrimination, where semantic editing increments provide stronger cues. 
Our method chiefly suppresses reference image redundancy, remaining competitive overall.

\textbf{FashionIQ:}
Table~\ref{tab:2} presents experimental results on the FashionIQ val set. 
(1) Our SDR-CIR achieves the best average R@10 and R@50 performance whether compared to textual inversion methods or training-free methods on the two backbones ViT-B/32, ViT-L/14.
In particular, on ViT-L/14 backbone, our method's average R@10 improves by 8.22\% compared to the current state-of-the-art method CoTMR. 
This demonstrates that our method not only excels in retrieval of object modifications, but also demonstrates great performance for attribute modifications.
(2) On ViT-G/14, average R@50 of our SDR-CIR is slightly below SEIZE, suggesting that larger backbones further amplify fine-grained editing cues of semantic editing increment in SEIZE. Notably, SDR-CIR is substantially more efficient. 
In summary, these results verify the strong generalization ability of SDR-CIR across different backbones and datasets, demonstrating its capability to mitigate semantic bias while maintaining high retrieval accuracy.
\begin{table*}[t]
\caption{Comparison results on the FashionIQ val set. The highest performance values are shown in \textbf{bold}, and the second-highest values are \underline{underlined}. "2S" indicates two-stage combination methods. "1S" indicates one-stage combination methods. "*" indicates experiments with GPT-4.1 as the MLLM and using CLIP weights from \cite{openclip_zenodo_5143773} without using in-context examples in the CoT prompts.}
\centering
\scalebox{0.8}{
\begin{tabular}{@{}c|ccc|ccccccccccc@{}}
\toprule
\multicolumn{4}{c|}{\textbf{FashionIQ→}} & \multicolumn{2}{c}{\textbf{Shirt}} & \multicolumn{2}{c}{\textbf{Dress}} & \multicolumn{2}{c}{\textbf{Toptee}} & \multicolumn{2}{c}{\textbf{Average}} \\ \midrule
Backbone & Method & Training-free & Type& R@10 & R@50 & R@10 & R@50 & R@10 & R@50 & R@10 & R@50 \\ \midrule
& SEARLE~\cite{baldrati2023zero} & × & - & 24.44 & 41.61 & 18.54 & 39.51 & 25.70 & 46.46 & 22.89 & 42.53 \\
\rowcolor{gray!20}
\multicolumn{1}{c|}{\cellcolor{white}{}} & CIReVL~\cite{karthik2023vision} & \ensuremath{\surd} & 2S& 28.36 & 47.84 & 25.29 & 46.36 & 31.21 & 53.85 & 28.29 & 49.35 \\
\rowcolor{gray!20}
\multicolumn{1}{c|}{\cellcolor{white}{}} & LDRE~\cite{yang2024ldre} & \ensuremath{\surd} & 2S& 27.38 & 46.27 & 19.97 & 41.84 & 27.07 & 48.78 & 24.81 & 45.63 \\
\rowcolor{gray!20}
\multicolumn{1}{c|}{\cellcolor{white}{}} & SEIZE~\cite{yang2024semantic} & \ensuremath{\surd} & 2S& 29.38 & 47.97 & 25.37 & 46.84 & 32.07 & 54.78 & 28.94 & 49.86 \\
 & OSrCIR*~\cite{tang2025reason} & \ensuremath{\surd} &  1S & \underline{31.31} & \underline{50.74} & 27.07 & 48.83 & 33.71 & 55.23 & 30.70 & 51.60 \\
 & CoTMR*~\cite{Sun_2025_ICCV} & \ensuremath{\surd} & 1S & 30.77 & 50.54 & \underline{29.45} & \underline{50.92} & \underline{34.93} & \underline{57.57} & \underline{31.72} & \underline{53.01} \\
 \multicolumn{1}{c|}{\multirow{-7}{*}{ViT-B/32}} & \textbf{SDR-CIR} & \ensuremath{\surd} & 1S & \textbf{36.41}& \textbf{57.02} & \textbf{36.84} & \textbf{58.85} & \textbf{43.14} & \textbf{64.71} & \textbf{38.80} & \textbf{60.19} \\ \midrule
 & Pic2Word~\cite{saito2023pic2word} & × & -& 26.20 & 43.60 & 20.00 & 40.20 & 27.90 & 47.40 & 24.70 & 43.70 \\
 & SEARLE~\cite{baldrati2023zero} & × & -& 26.89 & 45.58 & 20.48 & 43.13 & 29.32 & 49.97 & 25.56 & 46.23 \\
  & MLLM-I2W~\cite{bao2025mllm} & × & -& 27.30 & 46.50 & 29.90 & 48.60 & 33.80 & 55.20 & 30.30 & 50.10 \\
\rowcolor{gray!20}
\multicolumn{1}{c|}{\cellcolor{white}{}} & CIReVL~\cite{karthik2023vision} & \ensuremath{\surd} & 2S& 29.49 & 47.40 & 24.79 & 44.76 & 31.36 & 53.65 & 28.55 & 48.57 \\
\rowcolor{gray!20}
\multicolumn{1}{c|}{\cellcolor{white}{}} & LDRE~\cite{yang2024ldre} & \ensuremath{\surd} & 2S& 31.04 & 51.22 & 22.93 & 46.76 & 31.57 & 53.64 & 28.51 & 50.54 \\
\rowcolor{gray!20}
\multicolumn{1}{c|}{\cellcolor{white}{}} & SEIZE~\cite{yang2024semantic} & \ensuremath{\surd} & 2S& 33.04 & 53.22 & 30.93 & 50.76 & 35.57 & 58.64 & 33.18 & 54.21 \\
 & OSrCIR*~\cite{tang2025reason} & \ensuremath{\surd} &  1S & \underline{34.00} & 51.86 & 27.57 & 48.34 & 32.94 & 54.46 & 31.50 & 51.55 \\
 & CoTMR*~\cite{Sun_2025_ICCV} & \ensuremath{\surd} & 1S& 33.42 & \underline{52.31} & \underline{29.70} & \underline{50.42} & \underline{34.73} & \underline{57.47} & \underline{32.62} & \underline{53.40} \\
  \multicolumn{1}{c|}{\multirow{-9}{*}{ViT-L/14}} & \textbf{SDR-CIR} & \ensuremath{\surd} & 1S & \textbf{41.02} & \textbf{59.27} & \textbf{37.04} & \textbf{59.15} & \textbf{44.47} & \textbf{65.32} & \textbf{40.84} & \textbf{61.25} \\ \midrule
\rowcolor{gray!20}
\multicolumn{1}{c|}{\cellcolor{white}{}} & CIReVL~\cite{karthik2023vision} & \ensuremath{\surd} & 2S & 33.71 & 51.42 & 27.07 & 49.53 & 35.80 & 56.14 & 32.19 & 52.36 \\
\rowcolor{gray!20}
\multicolumn{1}{c|}{\cellcolor{white}{}} & LDRE~\cite{yang2024ldre} & \ensuremath{\surd} & 2S & 35.94 & 58.58 & 26.11 & 51.12 & 35.42 & 56.67 & 32.49 & 55.46 \\
\rowcolor{gray!20}
\multicolumn{1}{c|}{\cellcolor{white}{}} & SEIZE~\cite{yang2024semantic} & \ensuremath{\surd} & 2S & \underline{43.60} & \textbf{65.42} & \underline{39.61} & \underline{61.02} & \underline{45.94} & \textbf{71.12} & \underline{43.05} & \textbf{65.85} \\
 & OSrCIR*~\cite{tang2025reason} & \ensuremath{\surd} & 1S & 35.03 & 54.02 & 30.94 & 53.50 & 36.66 & 58.29 & 34.21 & 55.27 \\
 & CoTMR*~\cite{Sun_2025_ICCV} & \ensuremath{\surd} & 1S & 35.03 & 53.39 & 32.82 & 55.48 & 37.89 & 59.46 & 35.25 & 56.11 \\
  \multicolumn{1}{c|}{\multirow{-6}{*}{ViT-G/14}} & \textbf{SDR-CIR} & \ensuremath{\surd} & 1S & \textbf{44.55} & \underline{62.37} & \textbf{42.74} & \textbf{63.41} & \textbf{48.29} & \underline{69.71} & \textbf{45.19}  & \underline{65.16} \\ \bottomrule
\end{tabular}
}
\label{tab:2} 
\end{table*}
\subsection{Ablation Study}
We conduct ablation experiments on the CIRCO test set and FashionIQ val set, using ViT-L/14 as the backbone to verify the effects of different modules in our method.
The ablation study results are shown in Table~\ref{tab:3}.

\textbf{Effect of Selective CoT prompt.}
The ablation results of the CoT show that:
\textbf{(1) w/o CoT}: Without CoT reasoning. 
On the CIRCO dataset, removing CoT results in a significant decrease of 6.09\% in mAP@5.
On the FashionIQ dataset, removing CoT results in a 2.27\% decrease in average R@10 compared to the full SDR-CIR. 
This difference demonstrates the importance of CoT reasoning for understanding the reference image and the modification text. 
Furthermore, because images in FashionIQ are simpler and less noisy, removing CoT results in a smaller performance drop than on CIRCO. 
This suggests that CoT is more effective for datasets with more complex images. 
\textbf{(2) w/o Selective extraction}: 
On the CIRCO and FashionIQ datasets, comprehensive image understanding in CoT causes a 0.26\% drop in mAP@5 and a 0.33\% drop in R@10. 
These small but consistent declines indicate that CoT alone has limited ability to address redundancy.
\begin{table}[!t]
\caption{Ablation study on CIRCO test set and FashionIQ val set with ViT-L/14.}
    \scalebox{0.9}{
    \begin{tabular}{@{}lccccc@{}}
    \toprule
    \multirow{2}{*}{\textbf{Method}} & \multicolumn{3}{c}{\textbf{CIRCO}}& \multicolumn{2}{c}{\textbf{FashionIQ}} \\  & k=5 & k=10 & k=25 & k=10 & k=50 \\ \midrule
    SDR-CIR & 30.91 & 31.50 & 34.03 & 40.84 & 61.25 \\ \midrule
    \multicolumn{6}{@{}l}{\textbf{1. Chain-of-Thought}}\\
    \quad  w/o CoT & 24.82 & 25.61 & 27.79 & 38.57 & 59.26 \\
    \quad  w/o Selective extraction & 30.65&31.42 &33.82 & 40.51&60.15\\
    \multicolumn{6}{@{}l}{\textbf{2. Key modules of SDR-CIR}}\\
    \quad  only description & 26.61 & 27.54 & 30.01 & 32.26 & 52.61 \\
    \quad   +\textit{Anchor} & 28.56 & 29.48 & 31.39 & 38.41 & 59.01 \\
    \quad   +\textit{Debias} & 26.88 & 27.86 & 30.38 & 33.63 & 54.43 \\
    \quad   +SDR & 30.91 & 31.50 & 34.03 & 40.84 & 61.25 \\ \midrule
    \multicolumn{6}{@{}l}{\textbf{3. Different MLLMs}}\\
    \quad  Qwen2.5-VL-72B & 27.82 & 28.46 & 30.80 & 38.71 & 58.60 \\
    \quad  GPT-4omini & 30.67 & 31.33 & 33.66 & 39.26 & 59.85 \\
    \quad  GPT-4.1 & 30.91& 31.50 & 34.03 & 40.84 & 61.25\\ \bottomrule
    \end{tabular}
}
\label{tab:3}
\end{table}

\textbf{Effect of key steps in SDR.}
\textbf{(1) Description-only. } Only using target image descriptions for retrieval. 
\textbf{(2) +\textit{Anchor}. } 
\begin{table}[!t]
\caption{Impact of SDR on different one-stage methods with ViT-L/14}
\scalebox{0.8}{
\begin{tabular}{@{}cccccc@{}}
\toprule
 &  & \multicolumn{4}{c}{\textbf{CIRCO~(mAP@k)}} \\ 
\multirow{-2}{*}{\textbf{Method}} & \multirow{-2}{*}{\textbf{SDR}} & k=5 & k=10 & k=25 & k=50 \\ \midrule
 & × & 21.83 & 22.46 & 24.49 & 25.44 \\
 & \ensuremath{\surd} & 25.86 & 26.56 & 28.73 & 29.77 \\
\multirow{-3}{*}{OSrCIR~\cite{tang2025reason}} & $\triangle$& {\color[HTML]{32CB00} +4.03} & {\color[HTML]{32CB00} +4.10} & {\color[HTML]{32CB00} +4.24} & {\color[HTML]{32CB00} +4.33}   \\ \midrule
 & × & 26.52 & 27.13 & 29.51 & 30.56 \\
 & \ensuremath{\surd} & 30.12 & 30.74 & 33.20 & 34.28 \\
\multirow{-3}{*}{CoTMR~\cite{Sun_2025_ICCV}} & $\triangle$ & {\color[HTML]{32CB00} +3.60} & {\color[HTML]{32CB00} +3.61} & {\color[HTML]{32CB00} +3.69} & {\color[HTML]{32CB00} +3.72} \\ \midrule
 & × & 26.61 & 27.54 & 30.01 & 31.14 \\
 & \ensuremath{\surd} & 30.91 & 31.50 & 34.03 & 35.08 \\
\multirow{-3}{*}{Ours} & $\triangle$ & {\color[HTML]{32CB00} +4.30} & {\color[HTML]{32CB00} +3.96} & {\color[HTML]{32CB00} +4.02} & {\color[HTML]{32CB00} +3.94} \\ \bottomrule
\end{tabular}
}
\label{tab:4}
\end{table}
Add the \textit{Anchor} step based on description-only. Compared to description-only, mAP@5 of CIRCO improves by 1.95\%, and average Recall@10 of the FashionIQ improves by 6.15\%. 
This demonstrates that image features can stabilize useful visual semantics and recover cues that would otherwise be omitted. 
The larger gains on FashionIQ suggest a higher incidence of missing cues in this dataset, making the integration of image features particularly beneficial.
\textbf{(3) +\textit{Debias}.} Add the Debias process based on description-only without using the \textit{Anchor} step. 
Compared to description-only, CIRCO and FashionIQ achieve only slight improvements of 0.27\% and 1.37\%, respectively. 
This suggests that directly suppressing visual semantic contribution in descriptions may weaken useful visual information, limiting the benefit. 
\textbf{(4) +SDR. } SDR-CIR. 
Compared to description-only, CIRCO improves by +4.30\% and FashionIQ improves by 8.58\%. Compared to +\textit{Anchor}, both datasets show further increases of 2.35\% and 2.43\%. 
This demonstrates that debiasing is more effective after anchoring and that the two steps are complementary within the SDR module.

\textbf{Comparison of different MLLMs.}
We compare different MLLMs. GPT-4.1 achieves the best results. 
Compared to GPT-4.1, Qwen2.5-VL-72B is lower by 3.09\% on CIRCO and by 2.13\% on FashionIQ.
GPT-4o-mini is highly comparable, with only 0.24\% lower on CIRCO and 1.58\% lower on FashionIQ, while being more efficient. 
These results show that our method remains strong across different MLLMs.

\subsection{In-depth Analysis}

\textbf{Effect of the Semantic Debias Ranking on one-stage ZS-CIR methods.}
As shown in Table~\ref{tab:4}, we validate the effectiveness of the proposed Semantic Debias Ranking~(SDR) mechanism based on the ViT-L/14 backbone. 
We use several one-stage ZS-CIR methods as baselines and do not use in-context examples in the CoT prompt for a fair comparison. 
The results show that methods incorporating the SDR module achieve varying degrees of improvement. 
The OSrCIR and CoTMR methods achieve significant improvements of 4.03\% and 3.60\% in mAP@5, respectively, after using the SDR module. 
This demonstrates that the SDR mitigates the semantic bias that MLLMs struggles to resolve in one-stage methods, thus verifying its plug-and-play applicability and effectiveness.

\noindent\textbf{Effect of hyper-parameters $\alpha$ and $\beta$.}
\begin{figure}[tb]
    \includegraphics[width=\columnwidth]{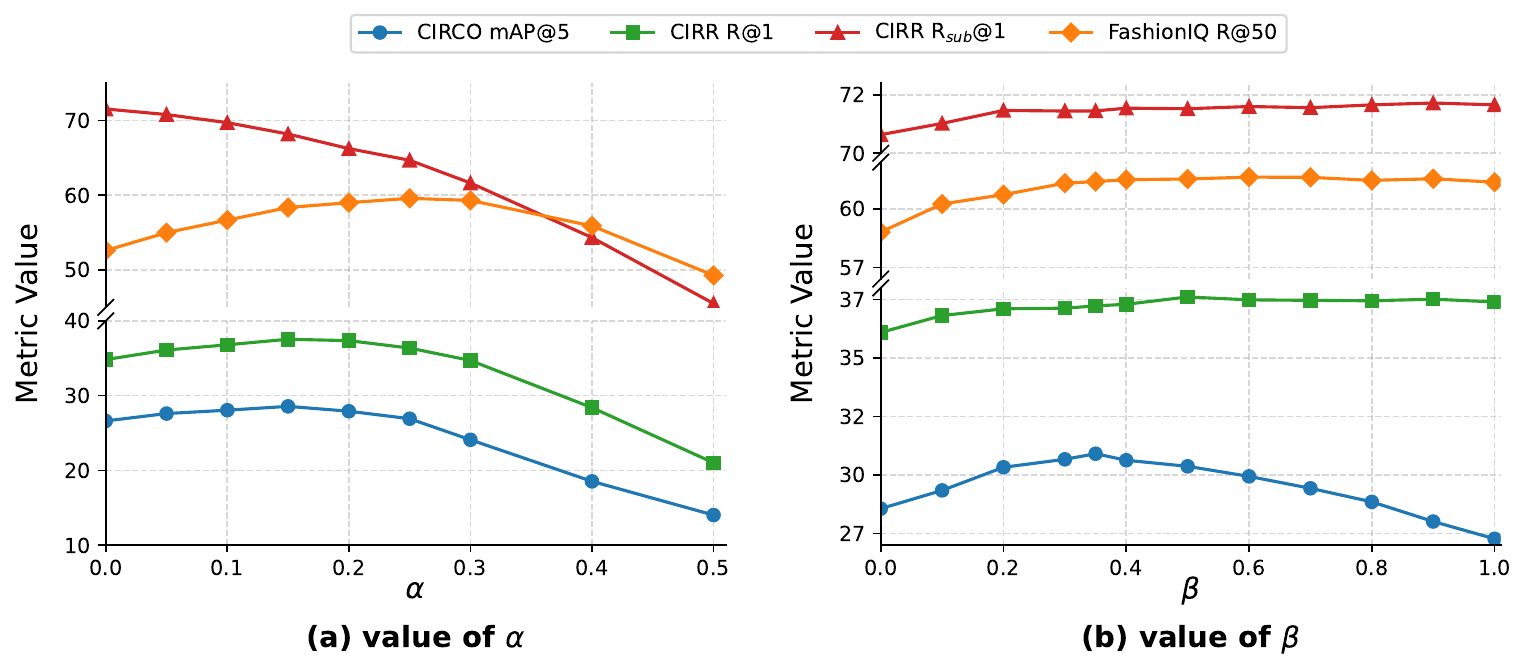} 
    \caption{Hyperparameter analysis of $\alpha$ and $\beta$ on CIRCO
test set, CIRR test set and FashionIQ val set. All experiments are performed with the ViT-L/14.}
    \label{fig:4}
\end{figure}
We analyze the hyperparameters in SDR-CIR via controlled experiments, as illustrated in Figure~\ref{fig:4}. 
As shown in Figure~\ref{fig:4}~(a), increasing $\alpha$ improves several metrics up to a moderate range, after which performance degrades.
FashionIQ continues to improve as $\alpha$ increases, stabilizing around $\alpha=0.2$. Notably, CIRR's R$_{\text{sub}}@1$ consistently shows a downward trend, indicating that excessively large $\alpha$ may amplify reference image noise, thereby weakening the subset retrieval accuracy. 
As shown in Figure~\ref{fig:4}~(b), as $\beta$ increases, overall performance of all datasets improves: CIRCO, FashionIQ reach their peaks at $\beta=0.35,0.4$; CIRR's R@1 and R$_{\text{sub}}@1$ are optimal at $\beta=0.5$.

\noindent\textbf{Efficiency Analysis.}
\begin{table}[tb]
\caption{Comparison with training-free methods on accuracy and efficiency metics with ViT-G/14. }
\scalebox{0.8}{%
\begin{tabular}{@{}ccccc@{}}
\toprule
\multirow{2}{*}{Method} & \multicolumn{2}{c}{Accuracy metic} & \multicolumn{2}{c}{Efficiency metic~(per query)} \\ \cmidrule(l){2-5} 
 & CIRCO~(k=1) & FashionIQ~(k=10) &  Infer time~(s)&  Calls~(times)\\ \midrule
LDRE~\cite{yang2024ldre} & 31.12 & 32.49 & 19.24 &15  \\
SEIZE~\cite{yang2024semantic} & 32.46 & 43.05 & 19.24 &15  \\
OSrCIR~\cite{tang2025reason} & 24.73 & 34.21 & 1.01 &1  \\
CoTMR~\cite{Sun_2025_ICCV} & 29.59 & 35.25 & 1.48 &2  \\
Ours & 33.05 & 45.19 & 0.37 & 1 \\ \bottomrule
\end{tabular}%
}
\label{tab:5}
\end{table}
As shown in Table~\ref{tab:5}, we compare the effectiveness and efficiency of common two-stage and one-stage training-free methods. 
Using the same backbone ViT-G/14, our method achieves the best performance among existing training-free methods, while significantly reducing the time. 
SEIZE, which has similar accuracy to our method, takes up to 19.24 seconds per query because it generates 15 descriptions and calls the LLM 15 times for each query. Therefore, our method achieves the best performance in both effectiveness and efficiency.


  

\subsection{Qualitative Results}
\begin{figure}[!t]
    \centering
 \includegraphics[width=\linewidth,trim=10 20 10 30, clip]{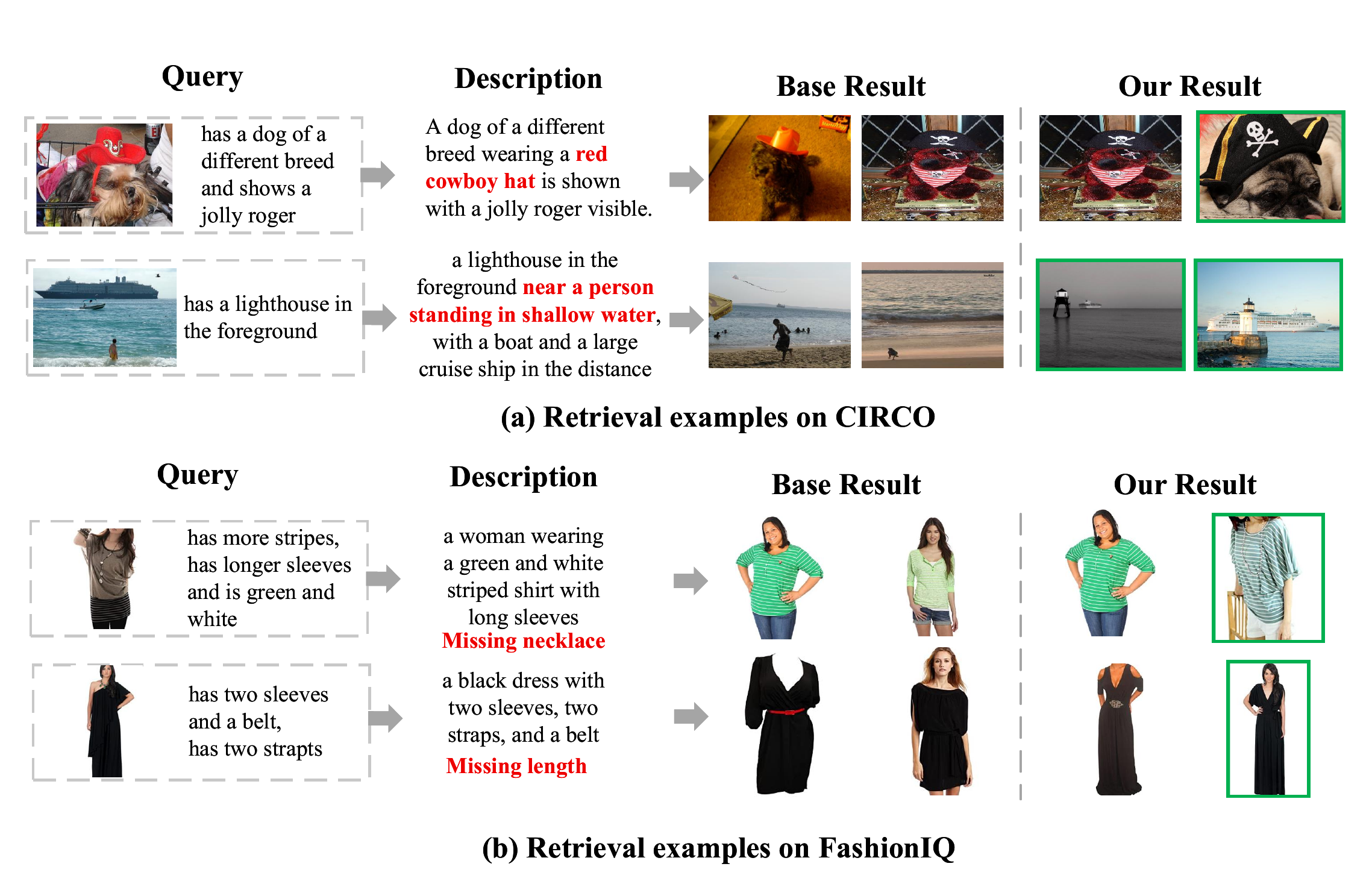}
\caption{Robustness to redundant and missing information on CIRCO and FashionIQ. Top-2 retrieval results of SDR-CIR and a description-only baseline (Base Result) are compared. Red text marks redundant or missing information; green boxes indicate targets.}
\label{fig:5}
\end{figure}

Figure~\ref{fig:5}~(a) illustrates examples of the redundant bias on CIRCO val set. 
In the first example, the description includes the redundant phrase “a red cowboy hat.” In the second, it includes “near a person standing in shallow water,” which is unrelated to the target image. 
Retrieval using only the description is steered by these details and retrieves incorrect images. 
With our Semantic Debias Ranking mechanism, the main information is preserved while the influence of redundant phrases is reduced, so the target image is retrieved.

Figure~\ref{fig:5}~(b) shows examples of the omission bias on FashionIQ val set. 
In these examples, the descriptions omit key cues related to the target image~(e.g., necklace, skirt length), causing description-only retrieval to fail to find the target image. 
Our method can supplement omitted elements in descriptions, which enables it to retrieve the target image in two cases.

\section{Conclusion}
In this paper, we propose SDR-CIR, a one-stage, training-free method for ZS-CIR that sequentially addresses the semantic bias introduced by reference images using Selective CoT and Semantic Debias Ranking. 
Selective CoT guides the MLLM to selectively extract the reference image content, thereby reducing the introduction of visual noise. 
Then, in the Semantic Debias Ranking, we obtain a robust composed query and explicitly model the visual semantic contribution of the reference image. 
Experiments on three benchmark CIR datasets demonstrate that our approach achieves state-of-the-art performance among one-stage methods, effectively mitigating the semantic bias.

\begin{acks}
\balance
This research is partially supported by the National Natural Science Foundation of China (Grant No. 62572365, 62271360), National Key Research and Development Program of China (Grant No. 2024YFF0907002)   
\end{acks}

\newpage
\bibliographystyle{ACM-Reference-Format}
\balance

\bibliography{reference} 

\newpage
\appendix

\section{Appendix}
\subsection{Selective CoT}
\begin{figure}[!h]
    \centering
    \includegraphics[width=\columnwidth]{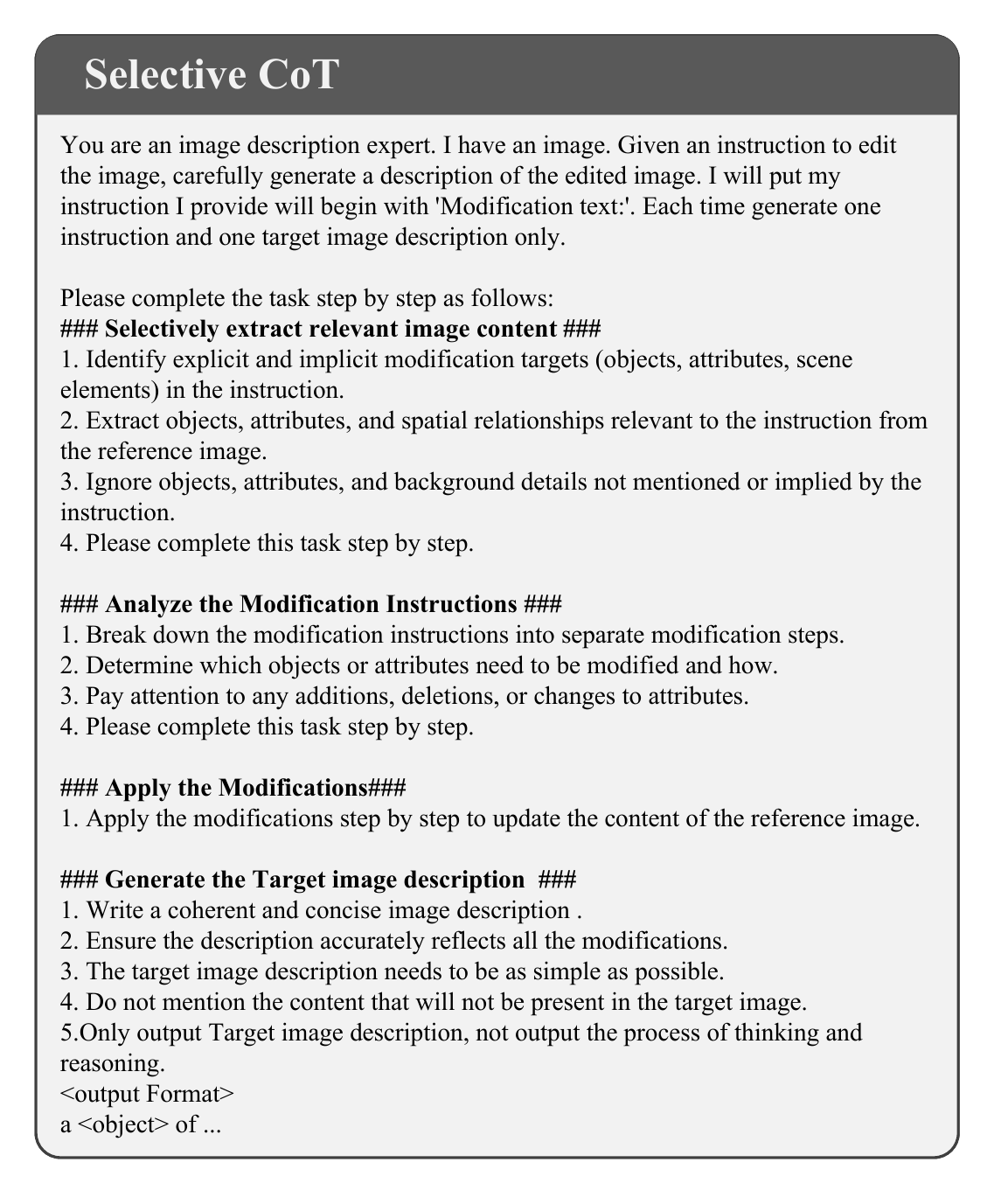}
    \caption{Our Selective CoT prompt. }
    \label{fig:6}
\end{figure}

Our Selective CoT is shown in the Figure~\ref{fig:6}.

\subsection{Failure Analysis}

\begin{figure}[!h]
    \centering
    \includegraphics[width=\columnwidth]{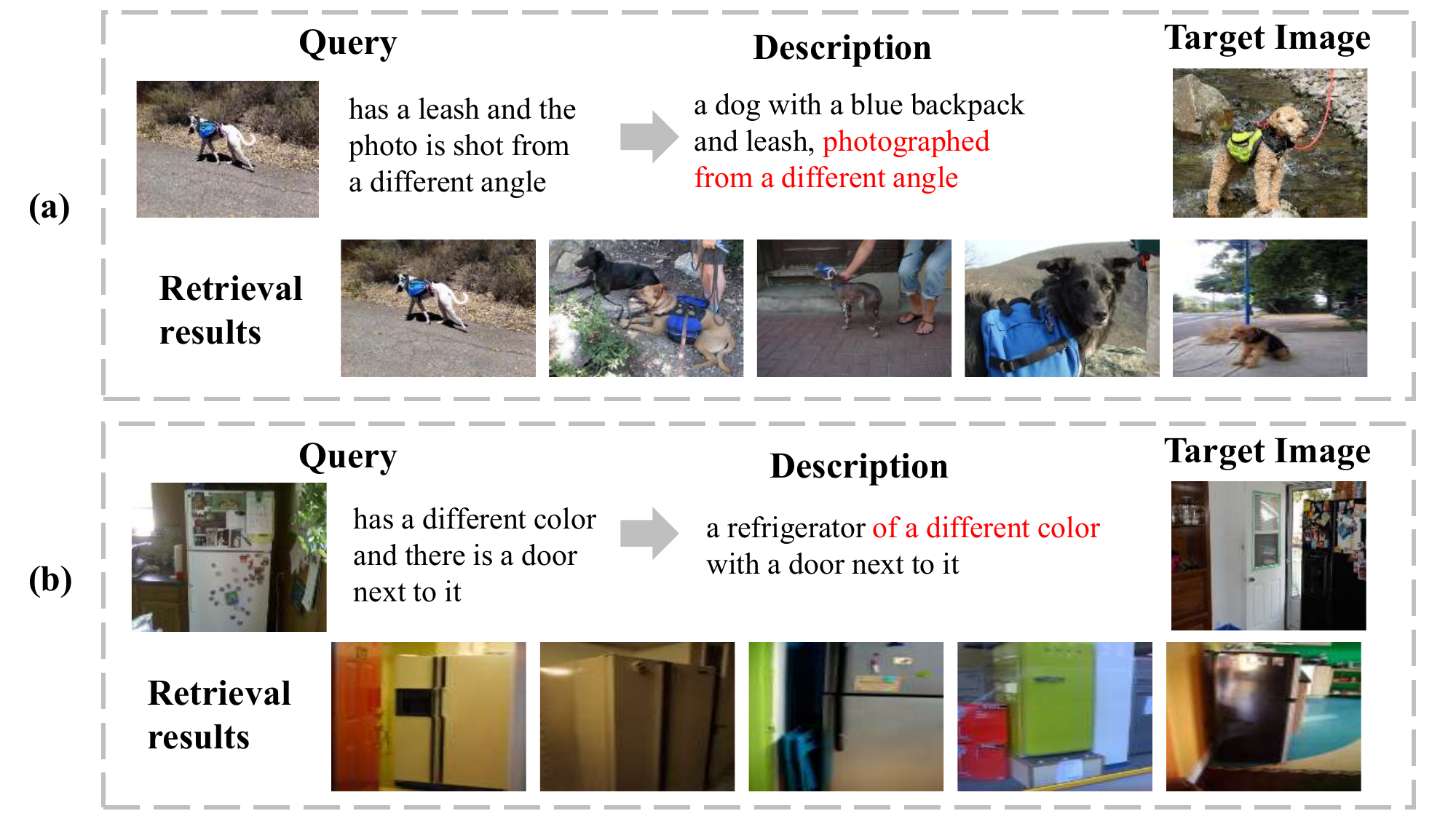} 
    \caption{Some failure cases of SDR-CIR.}
    \label{fig:7}
\end{figure}

To better understand the limitations of SDR-CIR, we present representative failure cases in Fig.~\ref{fig:7}. 
These examples highlight challenges under ambiguous editing intentions and complex visual scenes, suggesting directions for future improvements.

We find that when the modification text does not specify a particular change but only requires something different from the reference image, the MLLM often fails to infer the intended result.
For example, in Figure~\ref{fig:7}~(a), the modification text “the photo is shot from a different angle” only indicates that the angle of the target image should be different from the reference image and does not specify the exact change (e.g., top-down, low-angle, side view, or back view). 
In Figure~\ref{fig:7}~(b), the modification text "has a different color and there is a door next to it" indicates that the color of the target image should be different from the reference and but does not specify which color.
However, the target image has a fixed viewpoint/color, which leads to two issues
(1) the MLLM cannot determine the specific change due to the ambiguity of modification intent~("photographed from a different angle" in Figure~\ref{fig:7}~(a) description); 
(2) even if the MLLM determines the specific change, it may differ from the target image.
Consequently, these cases may not retrieve the target image.
These failures are mainly triggered by underspecified edits, where the modification does not provide sufficient constraints to uniquely determine the target, leading to plausible but non-discriminative descriptions.

\end{document}